\documentclass[pre,twocolumn,groupedaddress,showpacs]{revtex4}
\usepackage{graphicx}
\begin{document}
\title{Self-similar disk packings as model spatial scale-free networks}
\author{Jonathan P.~K.~Doye}
\email{jpkd1@cam.ac.uk}
\affiliation{University Chemical Laboratory, Lensfield Road, Cambridge CB2 1EW, United Kingdom}
\author{Claire P.~Massen}
\affiliation{University Chemical Laboratory, Lensfield Road, Cambridge CB2 1EW, United Kingdom}
\date{\today}
\begin{abstract}
The network of contacts in space-filling disk packings, such 
as the Apollonian packing, are examined. These networks
provide an interesting example of spatial scale-free networks, 
where the topology reflects the broad distribution of disk areas. 
A wide variety of topological and spatial properties 
of these systems are characterized. Their potential as models
for networks of connected minima on energy landscapes is discussed.
\end{abstract}
\maketitle

\section{Introduction}
Since the seminal paper of Watts and Strogatz on small-world networks 
\cite{Watts98},
there has been a surge of interest in complex networks, both
to characterize real-world networks and to generate 
network models to describe their properties 
\cite{Strogatz01,Albert02,Linked,Bornholdt03,Dorogovtsev03,Newman03a}.
The systems analysed in this way have spanned an
impressive range of fields, including astrophysics \cite{Hughes03}, 
geophysics \cite{Baiesi04}, information technology \cite{Albert99}, 
biochemistry \cite{Jeong00,Jeong01}, 
ecology \cite{Dunne02} and sociology \cite{Liljeros01}. 
Initially, the focus was on relatively basic topological properties of 
these networks, such as the average separation between nodes and 
the clustering coefficient to test whether they behaved like the 
Watts-Strogatz small-world networks \cite{Watts98}, or
the degree distribution to see if they could be classified as scale-free 
networks \cite{Barabasi99}.

As the field has progressed, however, the emphasis has shifted away
from these basic classifications 
to increasingly detailed characterization of the networks.
For example, on a topological level, there has been much recent interest 
in both the correlations \cite{Maslov02} and community structure 
\cite{Girvan02} within a network.

There has also been increasing interest
in how the medium in which a network is embedded influences the 
network properties. For spatial networks this can often lead to 
some kind of geographical localization \cite{Gastner04}.
For example, in social networks, acquaintances are more likely 
to share the same neighbourhood, and 
for the internet there is obviously a greater cost associated
with making longer physical connections \cite{Yook02}.
To model these kinds of effects there have been a number of studies
in which the preferential attachment rule that leads to scale-free 
networks \cite{Barabasi99} has been altered to include an additional 
distance dependence in the attachment 
probability \cite{Xulvi02,Manna02,Sen03,Barthelemy03}.
Typically, this leads to some crossover away from scale-free behaviour
when the distance constraint is sufficiently strong.

A different approach to understanding the interplay of geography and
topology has been to consider ways in which a scale-free
network can be embedded in Euclidean space 
\cite{Rozenfeld02,Warren02,benAvraham03,Herrmann03}. In most 
of these spatial scale-free networks, 
the nodes are distributed homogeneously in space 
\cite{Rozenfeld02,Warren02,benAvraham03}.
The heterogeneity that leads to the scale-free behaviour instead comes from
the node dependence of the range of interactions, i.e.\ high degree nodes
have connections to nodes that lie within a larger neighbourhood of the node.
The model of Herrmann {\it et al.}, however, shows the converse behaviour
\cite{Herrmann03}. Each node has the same interaction range; instead the 
scale-free behaviour is driven by an inhomogeneous density distribution with
high-degree nodes associated with regions of high node density.

Our interest in spatial scale-free networks comes from recent work 
characterizing the connectivity of the configuration space of atomic
clusters \cite{Doye02c,Doye04d}. Configuration space can be divided up
into basins of attraction surrounding each of the minima on the potential
energy surface of the clusters \cite{StillW84a}. 
This then allows a network description
of the potential energy surface where the nodes correspond to the minima, 
and two minima are linked if there is a transition state valley directly 
connecting them. All links are therefore between adjacent basins
of attraction. Intriguingly, this ``energy landscape'' network 
was found to be scale free. 
Since that initial study, the configuration space of some polypeptide
chains has also been found to have a scale-free 
connectivity \cite{Rao04}.

This scale-free behaviour cannot be explained by the usual preferential
attachment approach \cite{Barabasi99} because these networks are static, 
and are just determined by the potential for the system. 
Neither are the spatial scale-free models 
described above much help, because they are not contact networks between 
spatially adjacent regions. For example, if one were to associate each 
point in Euclidean space of these models with the nearest node 
the network of contacts between the resulting cells would not be scale-free.
Instead, the scale-free behaviour of these spatial networks arises precisely
because there are more long-range connections between non-adjacent nodes.

To try to further understand the energy landscape networks a different 
approach must be taken. In Ref.\ \cite{Doye02c} it was suggested that the 
scale-free behaviour might reflect differences in the basin areas, with
the deeper minima having large basins of attraction \cite{Doye98e} 
with many connections to the smaller basins surrounding them. For this to lead
to a scale-free topology, one would imagine that the basins
have to be arranged in some kind of hierarchical fashion with basins at 
each level being surrounded by successively smaller basins.

Space-filling disk packings, such as the Apollonian packing depicted
in Figure \ref{fig:Apollo}, have just such features. In this paper, 
we examine the contact networks for such packings to determine whether they
might provide a useful model for the energy landscape networks.
In the final stages of the preparation of this work, Andrade {\it et al.}
independently introduced the idea of Apollonian networks \cite{Andrade04}.
In that work, only a brief characterization of the topology of the
two-dimensional (2D) Apollonian packing was given, before the emphasis 
switched to the 
behaviour of dynamical processes on these networks. Here, we provide 
a much more detailed characterization of the topology of the 
2D network (Section \ref{sect:2D}), and also analyse the networks associated 
with other self-similar circle and hypersphere packings 
(Section \ref{sect:other}). Furthermore, as our aim is to
provide a model to help understand the energy landscape networks,
a particular emphasis is the relationship between the topological properties 
of the networks and the spatial properties
of the packings (Section \ref{sect:spatial}).

\section{Topological Properties}
\subsection{2D Apollonian networks}
\label{sect:2D}

\begin{figure}
\includegraphics[width=8.6cm]{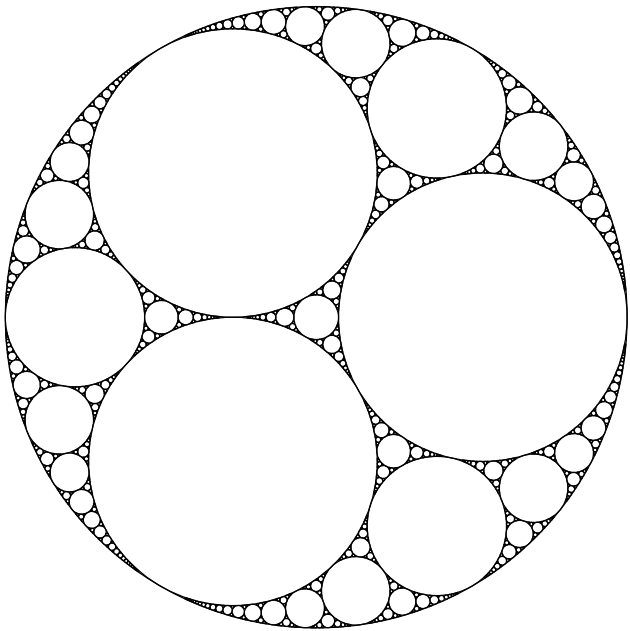}
\caption{An Apollonian packing of disks within a circle.}
\label{fig:Apollo}
\end{figure}

To produce an Apollonian packing,
we start with an initial array of touching disks, the interstices of which
are curvilinear triangles. In the first generation
disks are added inside each interstice in the initial configuration, 
such that these disks touch each of the disks bounding the curvilinear
triangles. The positions and radii of these disks can easily be calculated
using either the Soddy formula \cite{Soddy} or by applying circular inversions. 
Of course, these added disks cannot fill all of the space in 
the interstices, but instead give rise to three smaller interstices. 
In the second generation, further disks are added inside all of these 
new interstices, which again touch the surrounding disks. 
This process is then repeated for successive generations.
If we denote the number of generations by $t$, where $t=0$ corresponds to the 
initial configuration, as $t\rightarrow\infty$ the space-filling Apollonian
packing is obtained.

There are two initial configurations that are commonly used. 
The first was used for Figure 
\ref{fig:Apollo}, and has three mutually-touching disks all inside and in 
contact with a larger circle. This configuration has the useful feature that
all of the initial disks are inside a curvilinear triangle formed by the other
three circles, so that their initial environment is equivalent to that for any 
of the subsequent disks immediately after generation. 
This configuration is therefore more convenient when analytically deriving 
the properties of the packing, because for the most part the same formulae 
apply to the initial disks as to all other disks.
However, the spatial nature of the connections to the 
bounding circle are more ill-defined

\begin{figure}
\includegraphics[width=8.6cm]{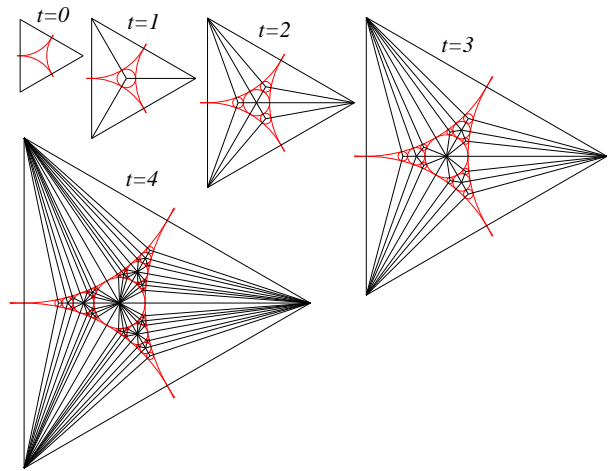}
\caption{(colour online) The development of the 2D Apollonian network
inside the interstice between three mutually touching disks, 
as the number of generations increases.
In each picture, the network is overlaid on the underlying packing.}
\label{fig:Apollo_net}
\end{figure}

The second common initial configuration is just to have three mutually 
touching disks as in Figure \ref{fig:Apollo_net}, and in subsequent generations 
to progressively fill 
the single curvilinear triangle in the initial configuration. The properties 
of the initial disks do not follow exactly the same trends as
the subsequent disks, but it has the advantage that none of the disks 
touch the interior boundary of another circle. The numerical results presented
will typically be for this initial triangular configuration.
However, we should emphasize that the networks associated with these
two initial configurations have the same topological properties, except for 
the initial disks.

Figure \ref{fig:Apollo_net} illustrates how the Apollonian packing can be
used as a basis for a network, where each disk is a node in 
the network and nodes are connected if the corresponding disks are in
contact. We shall call this contact network an Apollonian network. Of course,
it has an infinite number of nodes. However, we shall generally 
consider the network properties after a finite number of generations in the
development of the complete Apollonian network. From our analytical results
we can quickly obtain the properties of the complete network
by taking the limit of large $t$. 
However, the numerical results are necessarily limited
to finite-sized networks.

Figure \ref{fig:Apollo_net} also shows how the network evolves with the 
addition of new nodes at each generation.
For each new disk added, three new interstices in the packing
are created, that will be filled in the next generation.
Equivalently, for each new node added, three new triangles 
are created in the network, into which nodes will be inserted in
the next generation.
Therefore,
\begin{equation}
\Delta n_v(t)=n_v(t)-n_v(t-1)=3 \Delta n_v(t-1) \quad\ t>1, 
\end{equation}
where $n_v$ is the number of nodes in the network.

For the Apollonian packing of a circle, 
$n_v(0)=4$, and $\Delta n_v(1)=4$, if we treat the bounding circle
on the same footing as the other three initial disks.
It follows that
\begin{equation}
\Delta n_v(t)=4\,3^{t-1} \quad\ \hbox{and}\quad\
n_v(t)=2\left(3^t+1\right)
\label{eq:nv}
\end{equation}
The addition of each new node leads to three new edges. 
Therefore,
\begin{equation}
\Delta n_e(t)=n_e(t)-n_e(t-1)=3 \Delta n_v(t)=4\,3^t, 
\end{equation}
where $n_e$ is the number of edges in the network.
As $n_e(0)=6$
\begin{equation}
n_e(t)=2\, 3^{t+1}
\end{equation}
$p$, the fraction of all the possible pairs of nodes that are actually 
connected, is given by
\begin{eqnarray}
p&=&{2 n_e\over n_v(n_v-1)}= {3^{t+1}\over (3^t+1) (3^t+1/2)} \nonumber\\
&\approx& {1\over 3^{t-1}}\qquad\hbox{for large $t$}.
\label{eq:sparse}
\end{eqnarray}
Therefore, the Apollonian network becomes increasingly sparse 
as its size increases.
By contrast, the average degree tends to a limiting value:
\begin{eqnarray}
\langle k \rangle&=&{2 n_e\over n_v}={2\,3^{t+1}\over 3^t+1}\nonumber\\
           &\approx&6\qquad\hbox{for large $t$}.
\end{eqnarray}

Surrounding each node $i$ are $k_i$ empty (i.e.\ 
not enclosing any nodes) triangles. As 
new nodes are added at the centre of all these triangles, $k_i$ new 
connections will be created.
Therefore, at each step the degree of a node doubles, i.e.\ 
\begin{equation}
k_i(t+1)=2k_i(t).
\label{eq:kdouble}
\end{equation}
Such a rule expresses a preferential attachment \cite{Barabasi99}.
The number of new connections is linearly proportional to the degree.

If $t_{c,i}$ is the step at which a node $i$ is created, $k(t_{c,i})=3$ and
hence
\begin{equation}
k_i(t)=3\, 2^{t-t_{c,i}}.
\label{eq:kdiscrete}
\end{equation}
Therefore, $k$ can take a series of discrete values up to 
$k_{\rm max}=3\,2^t$. The fraction of the other nodes that the node
with maximum degree connects to is given by 
\begin{eqnarray}
{k_{\rm max}\over n_v-1}&=&{3\,2^t\over 2\,3^t+1}\nonumber\\
   &\approx& \left({2\over 3}\right)^{t-1}\qquad\hbox{for large $t$}
\end{eqnarray}
and is a decreasing fraction of the total as the size of the 
network increases.
It follows that the degree distribution is given by 
\begin{equation}
p(k)=\left\{\begin{array}{lc}
{\displaystyle{n_v(0)\over n_v(t)}={2\over 3^t+1} }
& \ \hbox{for}\ k=3\,2^{t}, t_c=0\\
{\displaystyle{\Delta n_v(t_c)\over n_v(t)}={2\,3^{t_c-1}\over 3^t+1} }
& \ \hbox{for}\ k=3\,2^{t-t_c}, t_c\ge 1\\
0 & \ \hbox{otherwise}\end{array} \right.
\end{equation}
and that the cumulative degree distribution is
\begin{equation}
p_{\rm cum}(k)={3^{t_c}+1\over 3^t+1}.
\end{equation}
Substituting for $t_c$ in this expression using $t_c=t-\log(k/3)/\log2$ gives
\begin{eqnarray}
p_{\rm cum}(k)&=&{3^{t}\left({k/3}\right)^{-\log3/\log2}+1\over 3^t+1}\nonumber\\
          &\approx&\left({k\over3}\right)^{\textstyle-{\log3\over \log2}}\qquad\hbox{for large $t$}.
\end{eqnarray}
For a continuous degree distribution $p(k)\sim k^{-\gamma}$, 
$p_{\rm cum}(k)\sim k^{1-\gamma}$. Therefore, the Apollonian 
network is scale free and the exponent of the degree 
distribution is
\begin{equation}
\gamma=1+{\log 3\over \log 2}=2.585,
\label{eq:gamma}
\end{equation}
as already noted in Ref.\ \onlinecite{Andrade04}.
Apollonian networks hence provide a new model for spatial scale-free networks.
Importantly, in contrast to other two-dimensional spatial scale-free networks
\cite{Xulvi02,Manna02,Sen03,Barthelemy03,Rozenfeld02,Warren02,benAvraham03,Herrmann03}, the 
Apollonian network can be embedded in a plane without any edges crossing.
In Aste {\it et al.}'s classification scheme, they hence have a genus of
zero \cite{Aste04}.

Another important property of the network is the clustering, which provides a 
measure of the local structure within the network.
The clustering coefficient of node $i$ is the probability that a pair of
neighbours of $i$ are themselves connected.
\begin{equation}
c_i(t)={2 n^{con}_i \over k_i(t) (k_i(t)-1)},
\label{eq:clocal}
\end{equation}
where $n^{con}_i$ is the number of connections between the neighbours of $i$.
At each stage a ring of new connections passing through all the nodes connected
to $i$ is generated. Therefore,
\begin{equation}
n^{con}_i(t)=\sum_{t'=t_{c,i}}^t k_i(t')=3\left(2^{t-t_{c,i}+1}-1\right),
\end{equation}
and 
\begin{eqnarray}
c_i(t)&=&{2 \left(2^{t-t_{c,i}+1}-1\right)\over 
2^{t-t_{c,i}} \left(3\,2^{t-t_{c,i}}-1\right)}\nonumber\\
      &\approx& {1\over 3\,2^{t-t_{c,i}-2}}={4\over k_i(t)}\qquad
      \hbox{for $t\gg t_{c,i}$}.
\end{eqnarray}
Therefore, the clustering coefficient of a vertex shows the same inverse
proportionality to the degree as has been obtained previously 
for other deterministic scale-free networks 
\cite{Dorogovtsev02,Ravasz03,Comellas04}.
This feature has been taken to be a signature of a hierarchical structure 
to the network \cite{Ravasz03,Barabasi04},
but recently has been shown to partially reflect disassortative 
correlations \cite{Soffer04}.
In the current networks this feature can also be 
interpreted in terms of spatial localization. For a low-degree node its
neighbourhood only occupies a small local region in the packing, and
thus would be expected to have strong clustering. By contrast, high-degree 
nodes have a more global character and are connected to well-separated
parts of the packing, and so have low clustering.

The clustering coefficient for the whole graph can be defined in two ways.
The first is a generalization of Eq.\ (\ref{eq:clocal}) to the whole graph,
and is the probability that any pair of nodes with a common neighbour
are themselves connected. Thus,
\begin{equation}
C_1={2 \sum_i n_i^{con}\over\sum_i k_i (k_i-1)}.
\label{eq:C1_def}
\end{equation}
The second definition of $C$ is as the average value of the local 
clustering coefficient, i.e.\
\begin{equation}
C_2={1\over n_v} \sum_i c_i .
\label{eq:C2_def}
\end{equation}
The difference between these two definitions is the relative weight
given to nodes with different degree. High degree nodes make a larger
contribution to $C_1$ because there will be more pairs of nodes that
have a high-degree node as a common neighbour, whereas all 
nodes contribute equally to $C_2$. Typically, $C_1<C_2$ 
because, as is the case here, higher degree nodes tend to have 
lower values of $c_i$.

Substituting in and rearranging gives
\begin{eqnarray}
C_1&=&{3^{t+1}-1 \over  12\left(2^{2t-1}-3^{t-1}\right)} \nonumber\\
 &\approx&{1\over 2}\left({3\over 4}\right)^t .
\end{eqnarray}
The clustering coefficient goes down as the size of the Apollonian network
increases. However when one compares $C$ to that for an Erd\H{o}s-Renyi 
random graph \cite{Erdos59,Erdos60} ($C_{\rm ER}=p$), one obtains
\begin{equation}
{C_1\over C_{\rm ER}}\approx {1\over 6}\left({3\over 2}\right)^{2t}.
\label{eq:C1vER}
\end{equation}
That is the Apollonian networks become increasingly more clustered than a 
random graph, as their size increases.
$C_2$ can be evaluated numerically. 
As shown in Ref.\ \onlinecite{Andrade04}, 
$C_2$ tends to a constant at large $t$ of value 0.828.

\begin{figure}
\includegraphics[width=8.6cm]{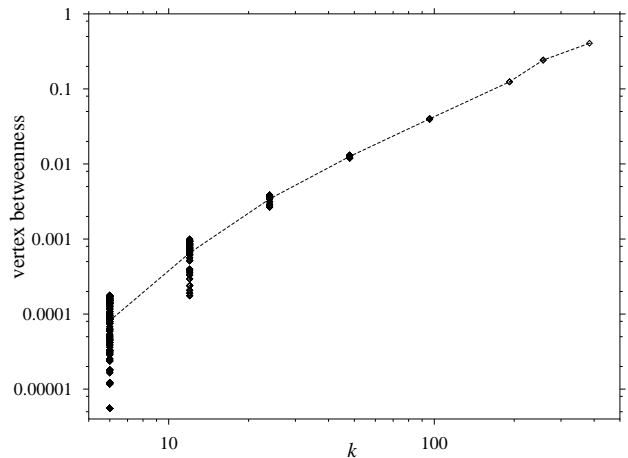}
\caption{The dependence of the vertex betweenness on degree
for the Apollonian network with $t=8$. There are points representing each disk
and the line connects the average values for a given $k$.}
\label{fig:between}
\end{figure}

In Ref.\ \onlinecite{Andrade04}, they also calculated the 
behaviour of $l_{\rm ave}$, the average
number of steps on the shortest path between any two nodes.
$l_{\rm ave}$ showed a small-world behaviour, scaling sub-logarithmically with 
network size \cite{Andrade04}. The important role played by the larger
disks in mediating these short paths is illustrated in Fig.\ \ref{fig:between}.
The vertex betweenness of a node is defined as the fraction of all the 
shortest paths that pass through that node. 
For $t=8$, 40\% of these paths pass 
through the central disk in the packing (Fig.\ \ref{fig:Apollo_net}). 
Furthermore, the dependence of the vertex betweenness on the degree
is not far from a power-law. 
This type of behaviour is common for scale-free networks \cite{Goh02}. 

Correlations in networks, particularly with respect to the degree, have
been the subject of increasing interest \cite{Maslov02}. This is partly
because the behaviour of models defined on such networks have been
often found to depend sensitively not only on the degree distribution, 
but also on how correlated the networks 
are \cite{Eguiluz02,Boguna03b,Vazquez03,Echenique04}.
For example, $k_{\rm nn}(k)$, the average degree of the neighbours of nodes 
with degree $k$, should be independent of $k$ for an uncorrelated network. 

We can calculate $k_{\rm nn}$ for the Apollonian network using 
Eq.\ (\ref{eq:kdouble}) to work out how many connections are 
made at a particular step to nodes with a particular degree. 
Except for the initial disks, no disks created in the same generation, i.e.\
with the same degree, will be connected. All connections to nodes with 
higher degree are made at the generation step, and then connections to 
lower degree nodes are made at each subsequent step. This leads to the
expression
\begin{eqnarray}
k_{\rm nn}(k)&=&{1\over \Delta n_v(t_c) k(t_c,t)}\Bigg( \nonumber\\
  &&\sum_{t'_c=0}^{t'_c=t_c-1} \Delta n_v(t'_c) k(t'_c,t_c-1)k(t'_c,t) +\nonumber\\
  &&\sum_{t'_c=t_c+1}^{t'_c=t} \Delta n_v(t_c) k(t_c,t'_c-1) k(t'_c,t)\Bigg)
\end{eqnarray}
for $k=3\,2^{t-t_c}$ and 
where $k(t_c,t)$ is the degree of a node at generation $t$ that
was created at generation $t_c$.
The first sum corresponds to the connections made to nodes with
higher degree (i.e.\ $t'_c<t_c$) when the node was created at $t_c$,
and the second sum to the connections made to the current lowest degree node
at each step $t'_c>t_c$. 
After substitution and evaluation of the sums, 
the above expression simplifies to
\begin{equation}
k_{\rm nn}(k)=9\left(4\over 3\right)^{t_c}-6 +{3\over 2}(t-t_c).
\label{eq:knn_tc}
\end{equation}
After the initial generation step $k_{\rm nn}$ for a node increases 
linearly with age.

\begin{figure}
\includegraphics[width=8.6cm]{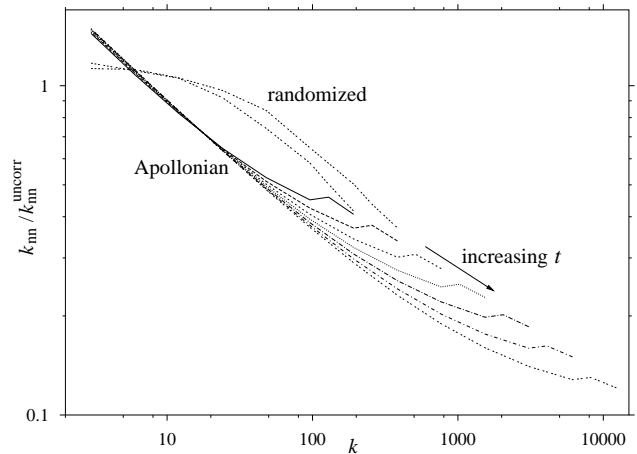}
\caption{Comparison of $k_{\rm nn}/k_{\rm nn}^{\rm uncorr}$ 
for the 2D Apollonian network to 
that for randomized versions of the networks. Lines corresponding
to different values of $t$ ($t=7$--15 and 7--8 for the Apollonian,
and randomized networks, respectively) have been plotted to
emphasize the universal form of this function for the Apollonian network.}
\label{fig:knn}
\end{figure}

Writing the above equation in terms of $k$ gives
\begin{equation}
k_{\rm nn}(k)=9\left({4\over3}\right)^t 
 \left({k\over 3}\right)^{\textstyle-{\log(4/3)\over\log 2}}-6 +{3\log(k/3)\over 2\log2}.
\end{equation}
$k_{\rm nn}(k)$ is roughly a power law function of $k$ with
exponent $-0.415$. 
The exponent is negative implying that the network is disassortative, 
i.e.\ nodes are more likely to be linked to nodes with dissimilar degree. 
When normalized by the expected value of $k$ for an uncorrelated network
\begin{equation}
k_{\rm nn}^{\rm uncorr}={\langle k^2\rangle \over\langle k\rangle}
                       =6\left({4\over 3}\right)^t - 3
\end{equation}
$k_{\rm nn}(k)$ has a universal form independent of network size for large $t$
and small $k$. 
Namely,
\begin{equation}
{k_{\rm nn}(k) \over k_{\rm nn}^{\rm uncorr}}\approx {3\over 2}
\left({k\over 3}\right)^{\textstyle-{\log(4/3)\over\log 2}}.
\label{eq:knn_norm}
\end{equation}
This is illustrated in Figure \ref{fig:knn}. The upward curvature away from
this power-law form at large $k$ is caused by the logarithmic term in $k$ in 
Equation (\ref{eq:knn_tc}).

It has been shown that disassortativity can often arise 
in networks where self-connections and multiple edges are excluded \cite{Park03}.
Therefore, $k_{\rm nn}$ was compared to that for random networks with 
the same degree distribution, which were prepared using the switching 
algorithm \cite{Maslov02,Milo04}. The randomized networks also show 
disassortativity, but to a somewhat lesser degree (Fig.\ \ref{fig:knn}).  
The additional disassortativity arises because connections 
to the nodes with the same degree cannot occur in the Apollonian network 
(except for the initial disks).

In particular, as the nodes with $k=3$ are only connected to 
higher degree nodes, $k_{\rm nn}(3)$ is significantly higher than 
that for the randomized graph (Figure \ref{fig:knn}). 
By contrast, $k_{\rm nn}$ for the rest of the network is 
lower than that for the randomized graphs.
This is also because of the lack of same-degree connections, as
this gives the higher degree nodes many more connections to the most numerous
$k=3$ nodes than for the randomized graphs. 

\begin{figure}
\includegraphics[width=8.6cm]{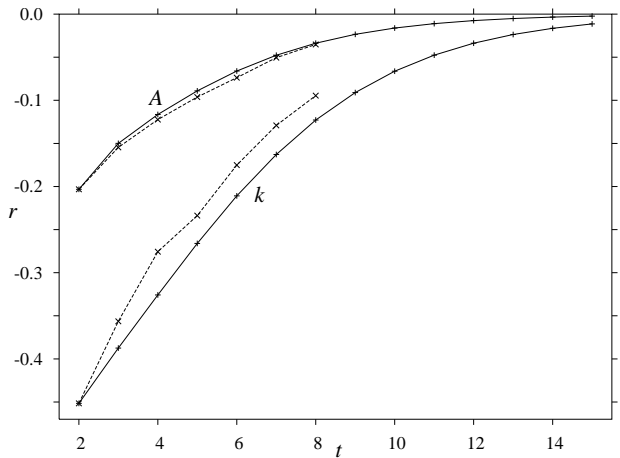}
\caption{Assortativity coefficients, $r_k$ and $r_A$, as a function of
the number of generations $t$. 
The solid lines are for the 2D Apollonian network, 
and the dashed lines for randomized versions of these networks 
with the same degree distribution.}
\label{fig:assort}
\end{figure}

An assortativity coefficient $r$ has been proposed that measures
the degree of (dis)assortativity of a property \cite{Newman02a},
and is defined as 
\begin{equation}
r={\langle s t\rangle_e - \langle s\rangle_e\langle t\rangle_e \over
    \langle s t\rangle_{e,assort} - \langle s\rangle_e \langle t\rangle_e }.
\label{eq:assort}
\end{equation}
where $s$ and $t$ correspond to the property of interest at
either end of an edge, $e$ denotes that the averages are over all edges
and {\it assort} that the average is for a perfectly assortative network.
$r$ is therefore a measure of the correlations in the property compared 
to that for a perfectly assortative network. 
Disassortative networks have $r<0$.

It follows that the assortativity coefficient for the degree is given by
\begin{equation}
r_k={\langle k\rangle\langle k^2 k_{\rm nn}(k)\rangle - 
    \langle k^2\rangle^2 \over
    \langle k\rangle \langle k^3\rangle - \langle k^2\rangle^2},
\end{equation}
Expressions for the quantities in the above equation can be relatively 
easily obtained from the degree distribution.
This gives 
\begin{eqnarray}
r_k&=&{3^t\left(2t-3+4\left({3\over 4}\right)^t\right)- 
      2^{2t}\left(2-\left({3\over 4}\right)^t\right)^2 \over
      {\displaystyle{3^t 2^{t}\over 5}}\left(6-\left({3\over 8}\right)^t\right)- 
      2^{2t}\left(2-\left({3\over 4}\right)^t\right)^2 } \nonumber\\
   &\approx&-{10\over3} \left({2\over 3}\right)^t\qquad\qquad\hbox{for large $t$.}
\label{eq:rk}
\end{eqnarray}
$r_k$ is always negative, indicating disassortativity. 
However, its magnitude goes to zero as the size of the network 
increases (Figure \ref{fig:assort}). 
This may seem surprising since $k_{\rm nn}$ always has a 
negative slope and has an effective functional form that is independent
of size (Eq.\ (\ref{eq:knn_norm})). However, 
the convergence of $r$ to zero is simply 
because the denominator corresponding to the correlations in a perfectly
assortative network scales more rapidly with size than the numerator,
i.e.\ as $6^t$ compared to $-4^t$ (Eq.\ (\ref{eq:rk})).

\begin{figure}
\includegraphics[width=8.6cm]{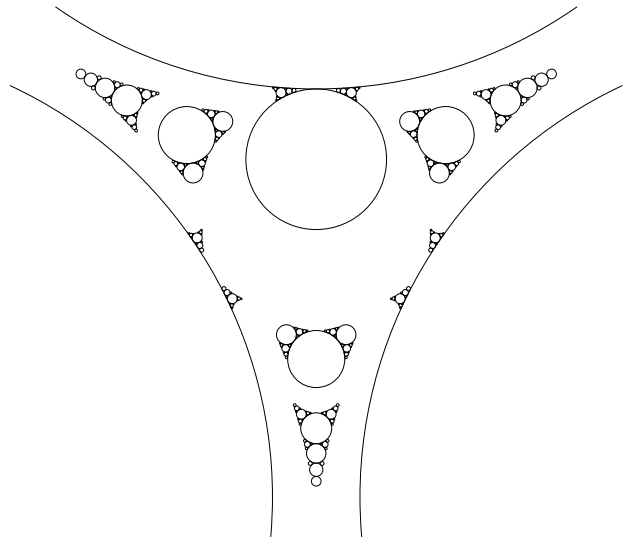}
\caption{The best division of the 2D Apollonian packing into
communities for $t=6$. The disks in the different communities have
been displaced with respect to each other for clarity. The modularity
$Q=0.5938$.}
\label{fig:community}
\end{figure}

We also looked at the community structure of this network using the
algorithm of Girvan and Newman \cite{Girvan02}.
It works by starting with the complete
network and at each step removing the edge that has the maximum edge 
betweenness, where this quantity is recalculated after the removal
of every edge and is defined as the fraction of all the shortest paths
that pass through an edge. If there is more than one edge with the same maximum
edge betweenness, they are all removed at the same step.
Thus the network is progressively divided into $n_v$
communities. To decide which division of the network represents the 
best choice, the modularity $Q$ is calculated at each step, and the division 
of the system with maximum $Q$ is considered to be the best \cite{Newman04b}.
$Q$ is defined as the fraction of edges that are within the communities 
compared to that expected for a random graph with the same degree 
distribution.

The best division of the packing into communities is shown 
in Figure \ref{fig:community} for $t=6$ and has $Q=0.5938$. This value is 
comparable to some of the higher values found for networks 
considered previously \cite{Newman04b,Newman04}. Interestingly, the Apollonian
network's combination of community structure and assortativity
is in contrast to most of the other networks with high $Q$
which tend to be also strongly assortative.
As expected the communities are spatially localized. As the algorithm
only used topological information, this result implies
that the spatial embedding of the network is clearly reflected in its 
topology.

\begin{figure}
\includegraphics[width=8.6cm]{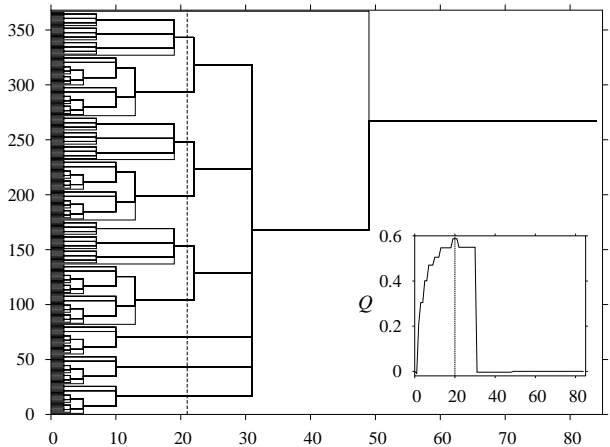}
\caption{A dendrogram showing the progressive division of the 
2D Apollonian packing into communities for $t=6$ as those
sets of edges with largest edge betweenness are removed.
At right no edges have been removed and there is a single community
of size $n_v$, whereas at left all edges have been removed and 
there are $n_v$ communities of size 1.
The number of steps in which sets of edges have been removed
increases linearly from right to left. In total 84 sets of edges were
removed. The three-fold symmetry of the packing is retained
throughout this process.  The vertical dashed line
indicates the set of communities with largest $Q$. 
The inset shows how $Q$ varies as the network is broken up into
communities.}
\label{fig:dendro}
\end{figure}

The algorithm for detecting communities described above cannot actually 
break the threefold symmetry of the packing. The best threefold symmetric
division has the central disk as its own community. However, by assigning
this disk to one of the adjacent communities, $Q$ improves from 0.5872 
to 0.5938.  We have also applied the faster algorithm described in 
Ref.\ \onlinecite{Newman04},
however slightly lower values of $Q$ were obtained.

It is interesting to examine how the network is progressively broken into
separate communities as more edges are removed. This can be represented by
a dendrogram as in Figure \ref{fig:dendro}, which shows the number of 
communities and their relationship at each step in the algorithm.
Every time two sets of disks become disconnected, their corresponding 
lines split. 
The first 53 sets of edges removed only break the packing into 
two communities; instead the effect is to make the contact matrix sparser.
Further removal of sets of edges then relatively quickly breaks the packing
into a large number of communities.

It is evident that $Q$ has quite a broad maximum as a function of 
the number of communities. $Q$ is greater then 0.3 
when there are between 6 and 88 communities.
Given the self-similar nature of the packing, one would not expect
there to be a strongly preferred size 
for the communities. Similarly, in the deterministic scale-free networks
the hierarchical modularity means that there is no clearly preferred
size for the modules at which the modularity is 
significantly enhanced \cite{Barabasi04}.

\subsection{Other self-similar packings}
\label{sect:other}

The 2D Apollonian network is only one example of a space-filling
self-similar packing.  In a similar way to the last section, a detailed 
characterization of the topology of contact networks associated with
other self-similar packings could be derived. 
However, here we do not wish to give such a comprehensive
account, but to illustrate how some of the key features of these networks, 
particularly the exponent of the degree distribution,
depend on the nature of the packing.

Firstly, we shall examine higher-dimensional Apollonian packings.
The initial configuration that is directly equivalent to Figure \ref{fig:Apollo}
is to have touching hyperspheres at the corners of a $d$-dimensional simplex 
that is enclosed within and touching a larger hypersphere.
The analysis of the last section is relatively easy to generalize to 
these cases.

As, now
$n_v(0)=\Delta n_v(1)=(d+2)$ and $\Delta n_v(t)=(d+1)\Delta n_v(t-1)$ for $t>1$,
it follows that 
\begin{equation}
\Delta n_v(t)=(d+2)(d+1)^{t-1}
\end{equation}
and
\begin{equation}
n_v(t)={(d+2)\over d}\left(\left(d+1\right)^t+d-1\right).
\label{eq:nv_d}
\end{equation}

The higher dimensional equivalent of Eq.\ (\ref{eq:kdouble}) is not so 
useful for calculating the degree distribution, for example for a 
3-dimensional Apollonian packing $k_i(t+1)=3k_i(t)-4$. Instead, an
alternative approach has to be used.
Each new neighbour of a node $i$ creates $d$ new $d$-simplices involving $i$.
In the next generation these $d$-simplices will be the sites for new
nodes that are also neighbours of i. Therefore,
\begin{eqnarray}
\Delta k_i(t)&=&k_i(t)-k_i(t-1) \nonumber\\
             &=&d\Delta k_i(t-1)\qquad{t>t_{c,i}+1}.
\end{eqnarray}
As $k_i(t_{c,i})=d+1$ and $\Delta k_i(t_{c,i}+1)=d+1$,
\begin{equation}
\Delta k_i(t)=(d+1)d^{t-t_{c,i}-1}
\end{equation}
and
\begin{equation}
k_i(t)={d+1\over d-1}\left(d^{t-t_{c,i}}+d-2\right).
\end{equation}
By an equivalent analysis to that for $d=2$ one can show that 
$p_{cum}(k)$ follows a power-law for large $t$ where 
\begin{equation}
\gamma=1+{\log(d+1)\over \log d}.
\end{equation}
Hence, the Apollonian networks associated with higher-dimensional 
packings are also scale-free networks.
The exponent $\gamma$ decreases as the dimension of the Apollonian
packing increases, tending to two in the limit of large $d$. This is 
noteworthy since the value of $\gamma$ 
can have significant effects on network properties \cite{Trusina04}.

By physical arguments it is easy to see that these higher-dimensional
Apollonian networks will have very similar topological 
properties to the two-dimensional case that we have studied in detail. 
The networks will again be disassortative with respect to 
degree because of the lack of connections between nodes with the 
same degree. The hierarchical structure and the more localized character 
of the connections involving low-degree nodes will lead to
a strong dependence of the clustering coefficient on degree.
This spatial localization will also lead to strong community structure.
Larger hyperspheres are also more likely to have a larger degree.

To illustrate how these conjectures can be backed up analytically,
here we derive a general 
expression for the local clustering coefficient. 
We first need to calculate the number of connections between the
neighbours of a node.
On generation a node is surrounded by a $d$-dimensional simplex,
and then at subsequent steps every new neighbour of a node
contributes $d$ new connections to $n_i^{con}$.
Hence,
\begin{eqnarray}
n_i^{con}(t)&=&{d(d+1)\over 2} + \sum_{t'=t_c+1}^t d \Delta k_i(t) \nonumber\\
&=&{d+1\over d-1} d^{t-t_c+1} + {d(d+1)(d-3)\over 2(d-1).} 
\end{eqnarray}
Substituting into Eq.\ \ref{eq:clocal} and taking limits gives:
\begin{equation}
c_i(t)\approx {2d \over k_i(t)}\qquad \hbox{for $t\gg t_{c,i}$}.
\end{equation}
Again, the local clustering coefficient is inversely proportional to
degree.

\begin{figure}
\includegraphics[width=8.6cm]{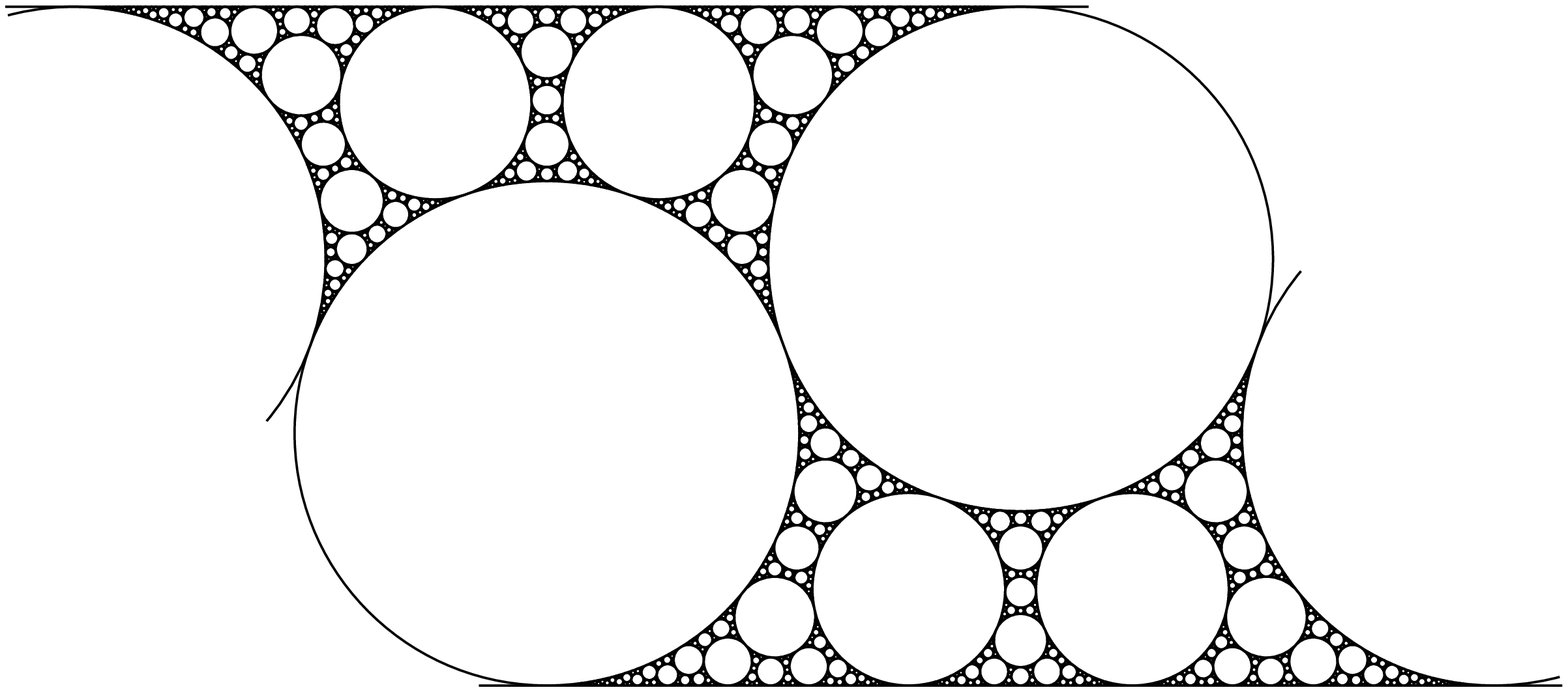}
\caption{A space-filling bearing 
between two half planes. 
The pattern can be repeated to the left and 
to the right {\em ad infinitum}.
In the numenclature of Refs.\ \onlinecite{Herrmann90,Manna91,Manna91b}
this packing is from Family 1 and has base loop size 4 and $n=m=2$. 
The half-planes at the top and bottom can be considered as disks with 
infinite radius.
}
\label{fig:sfb}
\end{figure}

In the Apollonian packing of disks, the smallest loops in the contact
network have size three, i.e.\ they are triangles.
However, space-filling packings of disks are possible, where the 
smallest loops in the contact network are polygons with more than three sides.
Examples, where the loops all have an even number of sides are of particular
interest, since they can act as space-filling bearings, where all the 
disks can rotate at the same time without slip \cite{Herrmann90,Oron00}.
An example of a space-filling bearings with `base loop size' 4 and
`$n=m$'.
is shown in Figure \ref{fig:sfb}.
The procedures to construct such packings are more complex than 
for Apollonian packings and are described in detail in Refs.\ 
\onlinecite{Herrmann90,Manna91,Manna91b}

\begin{figure}
\includegraphics[width=8.6cm]{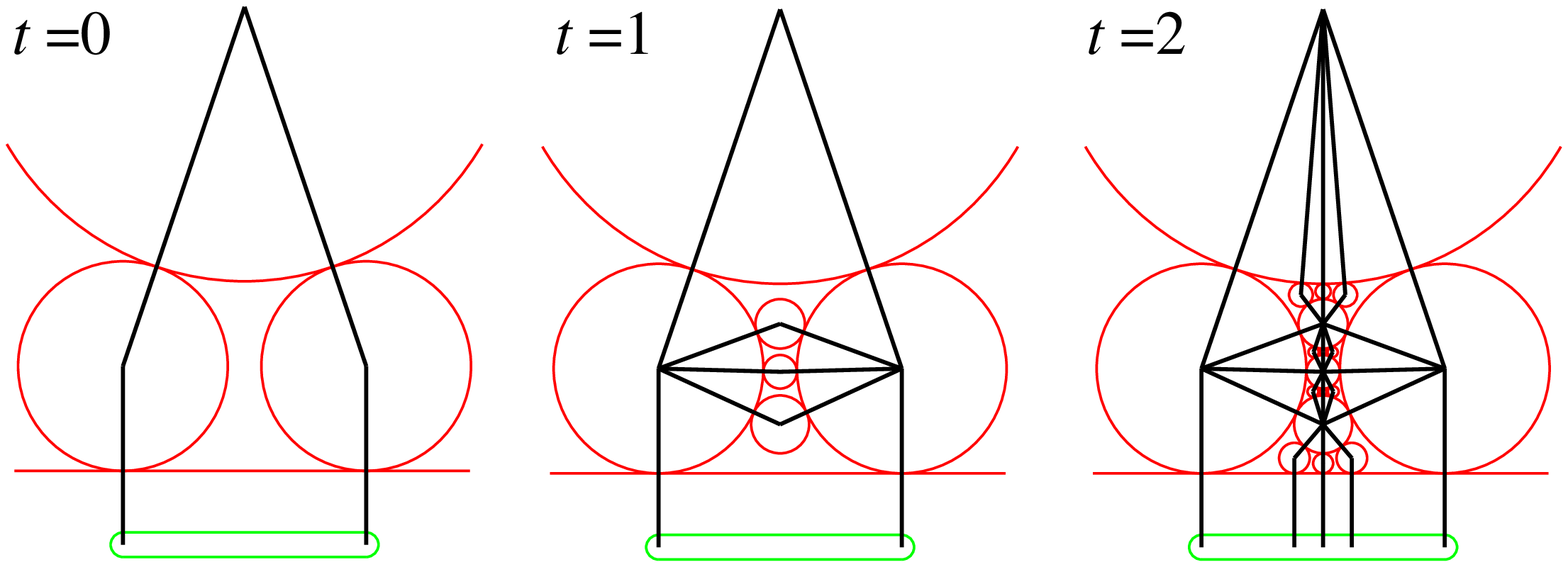}
\caption{(colour online) The development of the contact network for a subset 
of the packing in Fig.\ \ref{fig:sfb} initially consisting of four touching 
disks, one of which has infinite radius. The edges in the network 
connect the centres of the disks that touch. Those edges involving the 
infinitely large disk are parallel, but to emphasise that they do 
actually connect to the same node, a box has been added that enclosed 
the ends of these edges.}
\label{fig:sfb_net}
\end{figure}

Figure \ref{fig:sfb_net} illustrates how the contact network associated with
this packing develops for a subset initially consisting of four touching disks.
At each stage $m+1$ new nodes are produced for 
each empty quadrilateral, dividing the quadrilateral into a further 
$m+2$ new quadrilaterals to which new nodes will be added at the 
next generation.

Assuming that there is one quadrilateral initially ($t=0$), 
the number of empty quadrilaterals after step $t$ is $(m+2)^t$. Hence,
\begin{equation}
\Delta n_v(t)=(m+1)(m+2)^{t-1}
\end{equation}
and 
\begin{equation}
n'_v(t)=n_v(t)-n_v(0)=(m+2)^t-1.
\end{equation}
We exclude the initial boundary disks in the above because they have 
slightly different properties from the rest of the disks in the
iterative scheme. Besides, for sufficiently large $t$ their contribution
is negligible.

Aside from the first step after a node is created, a node only gains new 
connections at every other step, because the new connections are added
across alternating diagonals of the quadrilaterals.
\begin{equation}
k_i(t+2)=(m+2) k_i(t).
\end{equation}
Therefore, 
\begin{equation}
k_i(t)=2(m+2)^{(t-t_{c,i})/2}\quad\hbox{for $t-t_{c,i}$ even}.
\end{equation}
In the same way as before this leads to 
\begin{eqnarray}
p_{cum}(k)&=&{(m+2)^t(k/2)^{-2} -1 \over (m+2)^t -1} \nonumber\\
    &\approx&\left({k\over 2}\right)^{-2} \quad\hbox{for $t$ large}.
\end{eqnarray}
Hence $\gamma=-3$, independent of $m$. Unlike the Apollonian packings,
$\Delta n_v$ and $k_i$ increase by the same factor $m+2$, and so this
cancels. The power of two in the above equation arises, because this increase
in $k_i$ occurs only at every other step. The situations is more complicated
for the case $m\neq n$, because there are two sub-populations of disks, 
and the degree distribution of each sub-population obeys its own power law.

We should note that the contact networks 
for these space-filling bearings have no triangles, and so the clustering
coefficient defined by Eq.\ (\ref{eq:C1_def}) is zero. However,
generalized clustering coefficients probing higher-order loops have
been proposed. 
Clearly for these space-filling bearings the number of `squares' (loops
of length 4) will be significantly higher than for a random network,
although we have not sought to quantify this.

The results in this section illustrate that the contact
networks associated with other space-filling disk and hypersphere packings
are also scale-free, but that the exponent of the degree distribution 
is not a universal constant but depends on the nature 
of the packing. We could have also considered other examples, 
such as scale-free bearings with base loop size greater than 
four \cite{Oron00} and non-Apollonian packings of 
spheres \cite{Mahmoodi04,Mahmoodi04b}, and it is likely that these again 
show somewhat different behaviour.

\section{Spatial Properties}
\label{sect:spatial}

The 2D Apollonian packing is a well-known example of a fractal 
\cite{Mandelbrot},
and has many of the typical fractal properties. 
For example, inside every curvilinear triangle 
no matter how small the same pattern of disk packing reoccurs, i.e.\ it is 
self-similar. Similarly, the estimated total length of the circumferences 
of all the circles continues to increase as the resolution of the measurement
increase. One of the most important properties of such a packing is its 
fractal dimension. To understand this quantity we need to define more carefully
the set for which 
we wish to know the dimension. In the packing the disks are all
considered to be open, that is the set of points associated with a disk $D_i$
contains all the points inside the disk boundary, but not the boundary itself. 
The residual set $R$ is then the points that are not part of any of the
open disks in the packing, or more formally $R=U-\bigcup_{i=1}^\infty D_i$
where $U$ is the set that is being packed. $d_F$, the fractal dimension
of $R$, is the quantity of interest. 

$d_F$ must obey $1<d_F<2$. The upper bound is obvious,
because, by virtue of the space-filling nature of the 
Apollonian packing, $R$ must have zero area. The lower bound 
follows from the fact that $\sum_{i=1}^\infty r_i=\infty$, 
where $r_i$ is the radius of disk $D_i$; i.e.\ the total
length of the boundaries is infinite \cite{Wesler60}.
This result can most easily be visualized by projecting the boundaries of each
disk onto a diameter of the circle bounding the region that is being packed. 
Points on this diameter are projected onto infinitely often. These bounds imply
the dimension of $R$ must be fractional, and hence the packing is fractal. 

So far, no analytic formula for the value of the 
fractal dimension for the 2D Apollonian packing has been obtained, but 
instead its numerical value has beens estimated with increasing
precision \cite{Hirst67,Larman67,Boyd73b,Boyd82,Manna91,Thomas94}.
Its value is 1.3057. It has been suggested that the fractal dimension of
the Apollonian packing is the minimum for any space-filling disk 
packing \cite{Melzak66}, because at each step of the generating process
the disk with maximum possible radius is inserted into each curvilinear
triangle, thus 
maximizing the area of the region that must be outside of the residual set.
The fractal dimensions found for 2D space-filling bearings are 
consistent with this assertion \cite{Manna91,Oron00};
all have larger values than that for the Apollonian packing with
the largest found being 1.803.
In fact, as pointed out by Melzak, it is easy to generate a disk packing 
with dimension arbitrarily close to 2 \cite{Melzak66}.
If each disk in the Apollonian packing
in Figure \ref{fig:Apollo} is replaced by a suitably scaled image of the
whole Apollonian packing, a new packing is obtained with higher fractal
dimension. If this process is repeated {\it ad infinitum}, a disk packing with 
a fractal dimension of 2 is eventually obtained.

For space-filling packings of $d$-dimensional hyperspheres, there are similar
limits for the fractal dimension, namely $d-1<d_F<d$. The only calculations
have been for three dimensions. The Apollonian packing has 
$d_F=2.4739$ \cite{Borkovec94}, and the values for space-filling bearings
are again larger \cite{Mahmoodi04b}.

The fractal dimension is of particular interest here, because it 
provides a means to characterize the properties of the disk areas. 
Melzak introduced
the exponent of a packing, $d_r$, defining it as the minimum value of $s$
for which $\sum_{i=1}^\infty r_i^s$ no longer diverges. It was proved
by Boyd that for the Apollonian packing that $d_F=d_r$ \cite{Boyd73b}.
To examine the divergence properties of this sum we can replace the sum
by an integral, because the divergence is controlled by the disks with 
small radii, for which the distribution of radii is quasicontinuous.
As this distribution follows a power law, $p(r)\sim r^{-\beta}$ 
\cite{Melzak69},
we have 
\begin{eqnarray}
\sum_{i=1}^\infty r_i^s&\approx&\int^{r_{\rm max}}_0 r^s p(r) dr
\sim\int^{r_{\rm max}}_0 r^{s-\beta}\nonumber\\
&=&\left[{r^{s+1-\beta}\over s+1-\beta} \right]^{r_{\rm max}}_0\\
&=& \infty \qquad\hbox{if }s<\beta -1\nonumber
\end{eqnarray}
Hence, $d_F=d_r=\beta-1$. This allows the fractal dimension to be determined
from the exponent of the numerically obtained $p(r)$ \cite{Boyd82}.
It follows that the area distribution 
is given by $p(A)=A^{-(1+d_F/2)}$ for the 2D case and more generally
for packings of $d$-dimensional hyperspheres the volume 
distribution 
\begin{equation}
p(V)=V^{-(1+d_F/d)}\approx V^{-2}\qquad\hbox{for large $d$}.
\label{eq:pV}
\end{equation}
Given the bounds for $d_F$ the exponent must lie between 
$2-1/d$ and $2$.

\begin{figure}
\includegraphics[width=8.6cm]{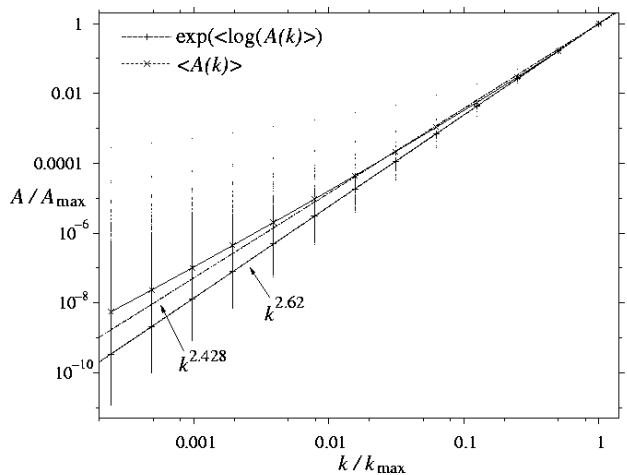}
\caption{The correlation of disk area with degree for the 2D Apollonian
packing (iterated to 15 generations). 
There is a dot representing each disk. The two lines with 
data points represent the average value of the area for a given degree,
and correspond to both the normal and logarithmic averages. In addition
lines corresponding to $(k/k_{\rm max})^{2.62}$ and $(k/k_{\rm max})^{2.428}$
have been plotted for comparison.}
\label{fig:Avk}
\end{figure}

One of the areas that is of most interest to us, and which is particularly
relevant to the energy landscape networks, is the connection between the 
spatial properties of the Apollonian packings and 
the topological properties of the Apollonian networks.
In Figure \ref{fig:Avk} we show the correlation between the disk area
and degree. As expected, the larger disks generally have a larger degree.
However, for a given $k$ there is a wide variety of disk areas. The
largest disks are associated with the crevices between the initial disks, 
whereas the smallest disks are obtained by following a spiral
pathway in the network where each disk along the path is connected to one
circle in each of the three previous generations.

More specifically, the logarithmic average of the disk area for a given $k$
closely follows a power-law. Assuming $A(k)\sim k^\alpha$ and using the 
identity $p(A) dA/dk =p(k)$ one can show that $\alpha=2(\gamma-1)/
(\beta-1)$. For the 2D Apollonian network this leads to the prediction 
$\alpha=2.428$. A line with this exponent is plotted for comparison in 
Fig.\ \ref{fig:Avk}, and broadly follows the average $\langle A(k)\rangle$.
By contrast, the average $\exp(\langle \log A(k)\rangle)$ has an exponent of
2.62.

Although the correlations associated with the degree are most commonly
studied, one can use Eq.\ (\ref{eq:assort}) to define an assortativity 
coefficient with respect to any property. 
Here, we examine the correlations in the areas
of touching disks. From the behaviour of $r_A$ in Fig.\ \ref{fig:assort} 
one can see that there is some slight disassortativity, but much weaker 
than for the degree and with little difference from that for the 
randomized graphs. However, if there was no variation in the
areas of the disks with a given $k$, (i.e.\ all the areas exactly
obeyed the power-law dependence of $k$ that characterizes the average)
then $r_A$ and $r_k$ would be very similar. 
Instead, because there is such a large scatter
around the average values---over seven decades for the $k=3$ disks in
Fig. \ref{fig:Avk}---the effect of the lack of connections between nodes with 
the same degree only weakly carries over to disks with similar areas. 

\begin{figure}
\includegraphics[width=8.6cm]{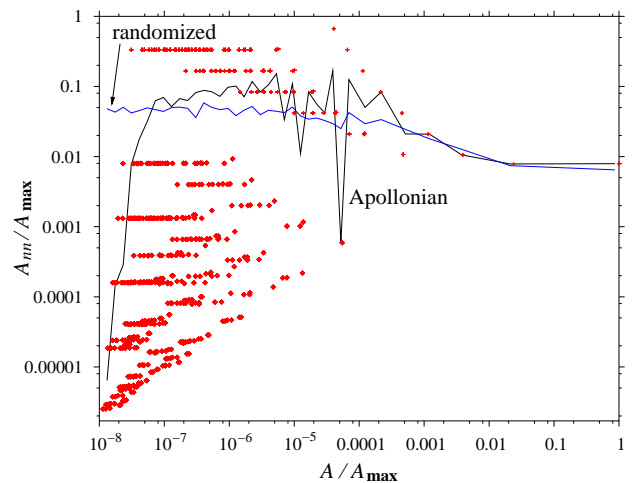}
\caption{(colour online) The average area of the disks that a touch a disk
against the area of that disk. The lines are binned averages for the Apollonian
networks with $t=8$ and a randomized version of that network.
There is a data point for each disk in the Apollonian packing. Those
represented by a cross are in contact with at least one of the three
initial disks, and those represented by a diamond are not. 
}
\label{fig:Ann}
\end{figure}

These effects can be examined in more detail by 
calculating $A_{\rm nn}$, the average area of the neighbours of a disk.
Weak disassortativity is evident over the majority of the 
range of areas, except for the smallest disks which show strong 
assortativity (Figure \ref{fig:Ann}). 
This latter effect is simply because the smaller disks in 
the last generation are connected to the smaller disks in the
previous generations, as with the spiral pathways mentioned above.
Interestingly, Fig.\ \ref{fig:Ann} clearly divides the disks into two sets
depending on whether they are in contact with one of the initial disks,
and this is the source of the large fluctuations in the average value of 
$A_{nn}$ at intermediate values of the disk area. 

\begin{figure}
\includegraphics[width=8.6cm]{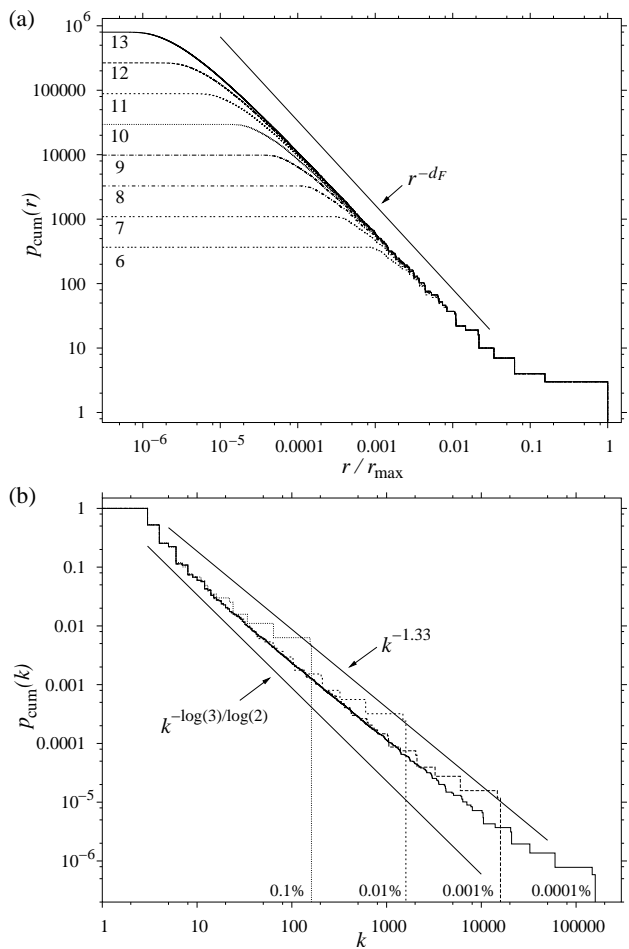}
\caption{(a) The cumulative size distribution of the disks for 
an Apollonian packing after a finite number of generations. 
The lines are labelled by their value of $t$, and an additional 
straight line with the exponent expected for the complete Apollonian 
packing has been added for comparison. 
(b) The cumulative degree distribution for a series of 
Apollonian network where only disks with radii at least $x$\% of that of 
the largest disk contribute.
The lines are labelled by their value of $x$, and additional 
straight lines have been added for comparison, one of which has the 
exponent expected for the complete Apollonian packing.}
\label{fig:finite}
\end{figure}

\section{Discussion}
\label{sect:compare}

Our main motivation for studying the properties
of the Apollonian networks is their potential to act
as a useful model for the energy landscape networks.
However, there is one major difference between the two systems. 
The Apollonian packings contain an infinite number of disks or hyperspheres, 
whereas configuration space is divided up into a finite number of basins,
albeit a number that is an exponentially increasing function of the number of 
atoms in the system \cite{StillW82,Doye02a}.

There are two ways of creating a finite network from the complete
Apollonian network. The first is to consider the network produced after a 
finite number of generations, and is the one we have mainly used so far. 
The second is to consider the network containing only disks that
are larger than a certain size. The wide distribution of areas for a given
$k$ in Fig. \ref{fig:Avk} indicates that their could
potentially be significant differences. We know that the first will have a 
scale-free degree distribution, and the second a power-law 
distribution of radii up to their respective cutoffs, but what about 
the other way round.

In Figure \ref{fig:finite}(a) the distribution of radii is shown after 
different numbers of generations. These distributions approximately
follow the expected power law for intermediate values of the radii, 
but this range becomes increasingly small as $t$ decreases. 
Furthermore, at small $r$ the lines curve away from this power law,
because the finite packings only contain a small fraction of the total
number of disks in the complete packing with that $r$.

The degree distributions for networks generated using a size cutoff 
are shown in Fig.\ \ref{fig:finite}(b).
The distributions still follow a power-law, and are actually smoother, 
since $k$ is no longer just restricted to the values
given by Eq.\ \ref{eq:kdiscrete}. However, the exponent is slightly 
smaller than predicted by Eq.\ (\ref{eq:gamma}). 
The effect of the size cutoff is to only 
include the larger disks from the later generations, which are in turn more
likely to be connected to the larger higher degree disks.
For example, for a radius cutoff at 0.0001\% of that of the largest disk,
the first disks below the size cutoff occurred in
the $13^{\rm th}$ generation, and the last disks included were in the 
$706^{\rm th}$ generation. 

Preliminary results for the basin area distributions for the small
clusters used to generate the energy landscape networks \cite{Massenunpub} 
look quite like Fig.\ \ref{fig:finite}(a) suggesting that Apollonian 
networks with a given number of generations are the more appropriate
finite version for comparison with these systems. Furthermore, there
are then some useful parallels between $t$
and $N$, the number of atoms in the cluster. For example, the number of minima
increases exponentially with $N$ and the number of disks/hyperspheres
in the Apollonian networks have a similar dependence on $t$ (Eqs.\ \ref{eq:nv}
and \ref{eq:nv_d}).
Similarly, as either $t$ or $N$ increase,
both types of networks become increasingly sparse (Eq.\ \ref{eq:sparse}), 
have a smaller absolute value for the clustering coefficient $C_1$ 
(Eq.\ \ref{eq:C1_def}), but a larger value relative to that for 
an Erd\H{o}s-Renyi random graph (Eq.\ \ref{eq:C1vER}) \cite{Doye02c,Doye04d}.

Other similarities between the two types of network include features that are
quite common for scale-free networks, such as the dependence of vertex
betweenness and local clustering coefficient on degree.
Both are also disassortative \cite{Doye04d}, however, 
there is greater community structure in the Apollonian networks 
\cite{Massenunpub}. There are also similar relationships between 
the topological and spatial properties, such as for the dependence
of disk or basin areas on $k$ \cite{Massenunpub}.

One of the interesting possibilities raised by the current study
is the signature of the scale-free topology of the Apollonian
network in the power-law behaviour of the disk areas.
Currently, mapping out the whole network of 
connections between minima on an energy landscape is only feasible 
for systems of very small size.  Neither are there 
methods available to construct a statistical representation of the whole
network from a finite sample.
Therefore, it is hard to test how generic is the scale-free behaviour 
observed for the clusters. However, the distribution of the hyperareas of 
the basins of attraction on a energy landscape is a static quantity that
could potentially be statistically sampled for a system of arbitrary size
\cite{Doye98e}. If this distribution exhibited a power law with exponent $-2$,
(Eq.\ (\ref{eq:pV})), it would strongly suggest that underlying this was
a scale-free energy landscape network, as for the Lennard-Jones clusters.
Preliminary calculations indicate that this is the case \cite{Massenunpub}. 

\section{Conclusions}
\label{sect:conc}
We have analysed the properties of the contact networks of
space-filling packings of disks and hyperspheres, focussing on the 
Apollonian packing of two-dimensional disks. 
Their topological properties include a scale-free degree distribution whose
exponent depends on the nature of the packing, high overall clustering, 
a local clustering coefficient that is inversely proportional to degree, 
disassortativity by degree and strong community structure.

These networks have many similarities to other deterministic scale-free 
networks introduced and analysed recently 
\cite{Barabasi01,Dorogovtsev02,Jung02,Ravasz03,Comellas04}, but with 
the additional feature that they have a well-defined spatial embedding.
For this reason, we have suggested that these packings provide a
useful model spatial scale-free network that may help to explain the
properties of energy landscapes and the associated scale-free network of
connected minima. In particular, the scale-free topology of the Apollonian
networks reflects the power-law distribution of disk sizes. Similarly,
configuration space can be divided up into basins of attraction that surround
the minima on the energy landscape. 
A similar power-law distribution for the hyperareas of these basins might 
thus provide an explanation for the pattern of connections between the minima.

\end{document}